\documentclass[preprintnumbers,amsmath,amssymbm,prd]{revtex4}
\usepackage{epsfig}
\usepackage{graphicx}
\usepackage{amssymb}

\begin{document}
\title{Quasi-bound state resonances of charged massive scalar
fields in the near-extremal Reissner-Nordstr\"om black-hole
spacetime}
\author{Shahar Hod}
\affiliation{The Ruppin Academic Center, Emeq Hefer 40250, Israel}
\affiliation{ }
\affiliation{The Hadassah Institute, Jerusalem
91010, Israel}
\date{\today}

\begin{abstract}
\ \ \ The quasi-bound states of charged massive scalar fields in the
near-extremal charged Reissner-Nordstr\"om black-hole spacetime are
studied {\it analytically}. These discrete resonant modes of the
composed black-hole-field system are characterized by the physically
motivated boundary condition of ingoing waves at the black-hole
horizon and exponentially decaying (bounded) radial eigenfunctions
at spatial infinity. Solving the Klein-Gordon wave equation for the
linearized scalar fields in the black-hole spacetime, we derive a
remarkably compact analytical formula for the complex frequency
spectrum which characterizes the quasi-bound state resonances of the
composed
Reissner-Nordstr\"om-black-hole-charged-massive-scalar-field system.
\end{abstract}
\bigskip
\maketitle


\section{Introduction}

The powerful no-hair theorems of Bekenstein and Mayo \cite{BekMay}
(see also \cite{Hodtp,Notewm,Hodrc,Herkr}) have revealed the fact
that, in asymptotically flat spacetimes, spherically symmetric
charged black holes cannot support external static matter
configurations made of charged massive scalar fields. It should be
emphasized, however, that these mathematically elegant no-hair
theorems \cite{BekMay} do not address the following physically
interesting question: how long does it take for a newly born charged
black hole to shed (that is, to swallow or to radiate to infinity)
the external charged fields?

Interestingly, recent numerical investigations \cite{Hercms} of the
Klein-Gordon wave equation for charged massive scalar fields in the
spherically symmetric Reissner-Nordstr\"om black-hole spacetime have
demonstrated explicitly that exponentially decaying quasi-bound
state charged matter configurations, which are characterized by
extremely long lifetimes \cite{Notelf}, can be supported in the
external spacetime region of the central charged black hole. These
external quasi-bound field configurations are characterized by the
physically motivated boundary condition of ingoing waves at the
outer black-hole horizon and exponentially decaying (bounded) radial
eigenfunctions at spatial infinity
\cite{Noteqnm,Hodch1,Hodch2,Konf,Ric}.

The main goal of the present paper is to explore the physical
properties of these quasi-bound state charged matter configurations
in the Reissner-Nordstr\"om black-hole spacetime. To this end, we
shall analyze the Klein-Gordon wave equation for the charged massive
scalar fields in the charged black-hole spacetime. Interestingly, as
we shall explicitly show below, in the extremal $Q/M\to1$ limit
\cite{Noteun,Notemq} of the central charged black hole, one can
derive a remarkably compact analytical formula for the discrete
frequency spectrum which characterizes the quasi-bound state
resonances of the composed
Reissner-Nordstr\"om-black-hole-charged-massive-scalar-field
configurations.

\section{Description of the system}

We consider a scalar field
$\Psi$ of mass $\mu$ and charge coupling constant $q$ \cite{Noteqm}
which is linearly coupled to a Reissner-Nordstr\"om black hole of mass $M$
and electric charged $Q$ \cite{Notegen}. The charged black-hole
spacetime is described by the spherically symmetric metric \cite{Chan}
\begin{equation}\label{Eq1}
ds^2=-f(r)dt^2+{1\over{f(r)}}dr^2+r^2(d\theta^2+\sin^2\theta
d\phi^2)\ ,
\end{equation}
where
\begin{equation}\label{Eq2}
f(r)=1-{{2M}\over{r}}+{{Q^2}\over{r^2}}\  .
\end{equation}
The horizon radii which characterize the charged
Reissner-Nordstr\"om black-hole spacetime,
\begin{equation}\label{Eq3}
r_{\pm}=M\pm(M^2-Q^2)^{1/2}\  ,
\end{equation}
are determined by the zeros of the dimensionless metric function
$f(r)$ \cite{Chan}.

The dynamics of the linearized charged massive scalar fields in the
curved black-hole spacetime is governed by the familiar Klein-Gordon
wave equation \cite{HodPirpam,Stro,HodCQG2,Hodch1,Hodch2,Konf,Ric}
\begin{equation}\label{Eq4}
[(\nabla^\nu-iqA^\nu)(\nabla_{\nu}-iqA_{\nu})-\mu^2]\Psi=0\  ,
\end{equation}
where $A_{\nu}=-\delta_{\nu}^{0}{Q/r}$ is the electromagnetic
potential of the Reissner-Nordstr\"om black hole. Substituting into
(\ref{Eq4}) the metric functions (\ref{Eq1}) and using the scalar
field expansion \cite{Noteom}
\begin{equation}\label{Eq5}
\Psi(t,r,\theta,\phi)=\int\sum_{lm}e^{im\phi}S_{lm}(\theta)R_{lm}(r;\omega)e^{-i\omega
t} d\omega\ ,
\end{equation}
one finds the characteristic radial differential equation \cite{HodPirpam,Stro,HodCQG2}
\begin{equation}\label{Eq6}
f(r){{d} \over{dr}}\Big[r^2f(r){{dR}\over{dr}}\Big]+UR=0\
\end{equation}
for the radial scalar eigenfunction $R(r)$, where \cite{Notell}
\begin{equation}\label{Eq7}
U=(\omega r-qQ)^2 -f(r)[\mu^2r^2+l(l+1)]\  .
\end{equation}

We shall be interested in resonant modes of the charged massive
scalar fields in the charged black-hole spacetime which are
characterized by the physically motivated boundary condition
\cite{Hercms,Notemas}:
\begin{equation}\label{Eq8}
R(r\to r_+)\sim e^{-i(\omega-\omega_{c})y}\
\end{equation}
of purely ingoing waves (as measured by a comoving observer) at the outer horizon of
the black hole, where
\begin{equation}\label{Eq9}
\omega_{\text{c}}={{qQ}\over{r_+}}\
\end{equation}
is the critical (marginal) frequency for the superradiant scattering
phenomenon of charged bosonic fields in the charged
Reissner-Nordstr\"om black-hole spacetime \cite{Bekc}, and the
radial coordinate $y$ is defined by the relation $dy/dr=1/f(r)$
\cite{Noteho}. In addition, the quasi-bound state resonances of the
linearized charged massive scalar fields that we shall analyze in
the present paper are characterized by the asymptotic boundary
condition \cite{Hercms,Notemas}
\begin{equation}\label{Eq10}
R(r\to\infty)\sim {{1}\over{r}}e^{-\sqrt{\mu^2-\omega^2}r}\
\end{equation}
of exponentially decaying radial eigenfunctions in the bounded regime
\begin{equation}\label{Eq11}
\omega^2<\mu^2\
\end{equation}
of small resonant frequencies.

The characteristic quasi-bound state resonances
$\{\omega(\mu,q,l,M,Q;n)\}$ \cite{Notenr} of the charged massive
scalar fields in the charged Reissner-Nordstr\"om black-hole
spacetime are determined by the set of equations (\ref{Eq6}),
(\ref{Eq8}), and (\ref{Eq10}).  Interestingly, as we shall
explicitly show in the next section, the complex resonant spectrum
which characterizes the quasi-bound state resonances of the composed
Reissner-Nordstr\"om-black-hole-charged-massive-scalar-field system
can be determined {\it analytically} in the near-extremal
$(r_+-r_-)/r_+\ll1$ regime.

\section{The quasi-bound state resonances of the charged massive
scalar fields in the charged Reissner-Nordstr\"om black-hole spacetime}

In the present section we shall study analytically the set of
equations (\ref{Eq6}), (\ref{Eq8}), and (\ref{Eq10}) which determine
the quasi-bound state resonances of the charged massive scalar
fields. In particular, below we shall derive a remarkably compact
analytical formula for the complex resonant spectrum which
characterizes the quasi-bound state massive scalar configurations in
the charged Reissner-Nordstr\"om black-hole spacetime.

It proves useful to define the dimensionless physical parameters
\cite{Teuk,Stro}
\begin{equation}\label{Eq12}
x\equiv {{r-r_+}\over {r_+}}\ \ \ ;\ \ \
\tau\equiv{{r_+-r_-}\over{r_+}}\ \ \ ;\ \ \ k\equiv 2\omega r_+-qQ\
\ \ ; \ \ \ \varpi\equiv{{\omega-\omega_{\text{c}}}\over{2\pi
T_{\text{BH}}}}\
\end{equation}
which characterize the composed
Reissner-Nordstr\"om-black-hole-charged-massive-scalar-field system,
where
\begin{equation}\label{Eq13}
T_{\text{BH}}={{r_+-r_-}\over{4\pi r^2_+}}\
\end{equation}
is the Bekenstein-Hawking temperature of the charged black hole. In
addition, we shall use the dimensionless black-hole-field parameter
\begin{equation}\label{Eq14}
\beta^2\equiv (l+1/2)^2+(\mu r_+)^2-k^2\  .
\end{equation}
Below we shall assume $0<\beta\in \mathbb{R}$ \cite{Notebi}.

Using the dimensionless parameters (\ref{Eq12}), one can express the
radial differential equation (\ref{Eq6}) in the from
\cite{Hodch1,Stro}
\begin{equation}\label{Eq15}
x(x+\tau){{d^2R}\over{dx^2}}+(2x+\tau){{dR}\over{dx}}+UR=0\  ,
\end{equation}
where the effective radial potential is given by
\begin{equation}\label{Eq16}
U=U(x;\mu,q,\omega,l,M,Q)={{(\omega
r_+x^2+kx+\varpi\tau/2)^2}\over{x(x+\tau)}}-l(l+1)-[\mu r_+(1+x)]^2\
.
\end{equation}
As shown in \cite{Hodch1}, for near-extremal black holes in the
$\tau\ll1$ regime, the characteristic differential equation
(\ref{Eq15}) is amenable to an analytical treatment in the two
asymptotic regions $x\ll1$ and $x\gg\tau\times\text{max}(1,\varpi)$
\cite{HodHod}. Interestingly, and most importantly for our analysis,
by performing a standard matching procedure of the two analytical
solutions [see Eqs. (\ref{Eq18}) and (\ref{Eq26}) below] in the
overlapping radial region $\tau\times\text{max}(1,\varpi)\ll x\ll 1$
\cite{Notematc}, one may determine the complex spectrum
$\{\omega(\mu,q,l,M,Q;n)\}$ of resonant frequencies which
characterize the quasi-bound state configurations of the charged
massive scalar fields in the charged Reissner-Nordstr\"om black-hole
spacetime.

We shall first analyze the radial equation (\ref{Eq15}) in the
near-horizon region
\begin{equation}\label{Eq17}
x\ll 1\  .
\end{equation}
In this spatial region one may approximate the effective radial
potential (\ref{Eq16}) by $U\to U_{\text{near}}\equiv
(kx+\varpi\tau/2)^2/[x(x+\tau)]-l(l+1)-(\mu r_+)^2$. One can express
the physically acceptable \cite{Notephr} near-horizon solution of
the radial differential equation (\ref{Eq15}) in the form
\cite{Hodch1,Abram,HodHod,Morse}
\begin{equation}\label{Eq18}
R(x)=x^{-i{{\varpi}\over{2}}}\Big({x\over
\tau}+1\Big)^{i{{\varpi}\over{2}}-ik}{_2F_1}({1\over
2}-\beta-ik,{1\over 2}+\beta-ik;1-i\varpi;-x/\tau)\  ,
\end{equation}
where ${_2F_1}(a,b;c;z)$ is the familiar hypergeometric function.

Using Eq. 15.3.7 of \cite{Abram} one may express (\ref{Eq18}) in the
form
\begin{eqnarray}\label{Eq19}
R(x)&=&x^{-i{{\varpi}\over{2}}}\Big({x\over
\tau}+1\Big)^{i{{\varpi}\over{2}}-ik}\Big[
{{\Gamma(1-i\varpi)\Gamma(-2\beta)}\over{\Gamma({1/2}-\beta-ik)\Gamma({1/2}-\beta+ik-i\varpi)}}
\Big({{x}\over{\tau}}\Big)^{-1/2-\beta+ik} \nonumber \\&& \times
{_2F_1}({1\over 2}+\beta-ik,{1\over
2}+\beta-ik+i\varpi;1+2\beta;-\tau/x)+(\beta\to -\beta)\Big]\ ,
\end{eqnarray}
where the notation $(\beta\to -\beta)$ means ``replace $\beta$ by
$-\beta$ in the preceding term". Substituting into (\ref{Eq19}) the
asymptotic expression (see Eq. 15.1.1 of \cite{Abram})
\begin{equation}\label{Eq20}
_2F_1(a,b;c;z)\to 1\ \ \ \text{for}\ \ \ {{ab}\over{c}}\cdot z\to 0\
,
\end{equation}
one obtains \cite{Hodch1}
\begin{eqnarray}\label{Eq21}
R(x)={{\Gamma(1-i\varpi)\Gamma(-2\beta)\tau^{1/2+\beta-i\varpi/2}}\over{\Gamma({1/2}-\beta-ik)
\Gamma({1/2}-\beta+ik-i\varpi)}}x^{-{{1}\over{2}}-\beta} +(\beta\to
-\beta)\
\end{eqnarray}
in the intermediate spatial region
\begin{equation}\label{Eq22}
\tau\times\text{max}(1,\varpi)\ll x\ll 1\  .
\end{equation}

We shall next analyze the radial scalar equation (\ref{Eq15}) in the
far-region
\begin{equation}\label{Eq23}
x\gg \tau\times\text{max}(1,\varpi)\  .
\end{equation}
In this spatial region one may approximate (\ref{Eq15}) by
\cite{Hodch1}
\begin{equation}\label{Eq24}
x^2{{d^2R}\over{dx^2}}+2x{{dR}\over{dx}}+U_{\text{far}}R=0\  ,
\end{equation}
where the effective far-region potential is given by $U\to
U_{\text{{far}}}=(\omega r_+x+k)^2-l(l+1)-[\mu r_+(1+x)]^2$.
Defining the dimensionless variables
\begin{equation}\label{Eq25}
\epsilon\equiv \sqrt{\mu^2-\omega^2}r_+\ \ \ \ ; \ \ \ \
\kappa\equiv {{\omega kr_+-(\mu r_+)^2}\over{\epsilon}}\  ,
\end{equation}
one can express the far-region solution of radial differential
equation (\ref{Eq24}) in the form \cite{Hodch1,Abram,HodHod,Morse}
\begin{equation}\label{Eq26}
R(x)=N_1\times(2\epsilon)^{{1\over 2}-\beta}x^{-{1\over
2}-\beta}e^{-\epsilon x}{_1F_1}({1\over
2}-\beta-\kappa,1-2\beta,2\epsilon x)+N_2\times(\beta\to -\beta)\ ,
\end{equation}
where ${_1F_1}(a,b,z)$ is the familiar confluent hypergeometric
function and $\{N_1,N_2\}$ are normalization constants.

Substituting into (\ref{Eq26}) the asymptotic expression (see Eq.
13.1.2 of \cite{Abram})
\begin{equation}\label{Eq27}
_1F_1(a,b,z)\to 1\ \ \ \text{for}\ \ \ {{a}\over{b}}\cdot z\to 0\  ,
\end{equation}
one obtains \cite{Hodch1}
\begin{equation}\label{Eq28}
R(x)=N_1\times(2\epsilon)^{{1\over 2}-\beta}x^{-{1\over
2}-\beta}+N_2\times(\beta\to -\beta)\
\end{equation}
in the intermediate spatial region
\begin{equation}\label{Eq29}
\tau\times\text{max}(1,\varpi)\ll x\ll 1\  .
\end{equation}

Matching the analytically derived radial solutions (\ref{Eq21}) and
(\ref{Eq28}), which are both valid in the overlap region
$\tau\times\text{max}(1,\varpi)\ll x\ll 1$, one finds the
normalization constants \cite{Hodch1,Noteesp}
\begin{equation}\label{Eq30}
N_1(\beta)={{\Gamma(1-i\varpi)\Gamma(-2\beta)}\over{\Gamma({1\over
2}-\beta-ik)\Gamma({1\over 2}-\beta+ik-i\varpi)}}\tau^{{1\over
2}+\beta-i{{\varpi}\over{2}}}(2\epsilon)^{-{1\over 2}+\beta}\ \ \ \
{\text{and}} \ \ \ \ N_2(\beta)=N_1(-\beta)\
\end{equation}
of the radial solution (\ref{Eq26}). In addition, using Eq. 13.5.1
of \cite{Abram}, one finds the asymptotic spatial behavior
\begin{eqnarray}\label{Eq31}
R(x\to\infty)&\to&
\Big[N_1\times(2\epsilon)^{\kappa}{{\Gamma(1-2\beta)}\over{\Gamma({1\over
2}-\beta+\kappa)}}x^{-1+\kappa}(-1)^{-{1\over
2}+\beta+\kappa}+N_2\times(\beta\to -\beta)\Big]e^{-\epsilon x}
\nonumber \\&& +
\Big[N_1\times(2\epsilon)^{-\kappa}{{\Gamma(1-2\beta)}\over{\Gamma({1\over
2}-\beta-\kappa)}}x^{-1-\kappa}+N_2\times(\beta\to
-\beta)\Big]e^{\epsilon x}\
\end{eqnarray}
for the radial eigenfunction (\ref{Eq26}). Remembering that the
quasi-bound configurations of the charged massive scalar fields in
the charged Reissner-Nordstr\"om black-hole spacetime are
characterized by exponentially decaying radial eigenfunctions at
spatial infinity [see Eqs. (\ref{Eq10}) and (\ref{Eq11})], one
arrives at the important conclusion that the coefficient of the
exploding exponent $e^{\epsilon x}$ in the asymptotic spatial
expression (\ref{Eq31}) should vanish:
\begin{eqnarray}\label{Eq32}
N_1\times(2\epsilon)^{-\kappa}{{\Gamma(1-2\beta)}\over{\Gamma({1\over
2}-\beta-\kappa)}}x^{-1-\kappa}+N_2\times(\beta\to -\beta)=0\  .
\end{eqnarray}
Finally, substituting the normalization constants (\ref{Eq30}) into
(\ref{Eq32}), one obtains the (rather cumbersome) resonance equation
\cite{Hodch1}
\begin{equation}\label{Eq33}
\Big[{{\Gamma(2\beta)}\over{\Gamma(-2\beta)}}\Big]^2{{\Gamma({1\over
2}-\beta-ik)\Gamma({1\over 2}-\beta-\kappa)\Gamma({1\over
2}-\beta+ik-i\varpi)}\over{\Gamma({1\over 2}+\beta-ik)\Gamma({1\over
2}+\beta-\kappa)\Gamma({1\over
2}+\beta+ik-i\varpi)}}=\big(2\epsilon\tau\big)^{2\beta}\
\end{equation}
for the complex frequency spectrum $\{\omega(\mu,q,l,M,Q;n)\}$ which
characterizes the quasi-bound state resonances of the linearized
charged massive scalar fields in the near-extremal charged
Reissner-Nordstr\"om black-hole spacetime.

The assumption $0<\beta\in \mathbb{R}$ \cite{Notebi} implies the
strong inequality
\begin{equation}\label{Eq34}
\delta\equiv\big(2\epsilon\tau\big)^{2\beta}\ll1\ \ \ \ \text{for}\ \ \ \
\tau\to0\
\end{equation}
in the low-temperature $\tau\to0$ limit. In this regime of
near-extremal ($\tau\to0$) black holes, the discrete resonant
solutions of Eq. (\ref{Eq33}) can be expressed in the compact form
\begin{equation}\label{Eq35}
\varpi_n\equiv k-i(n+{1\over2}+\beta-\eta\cdot\delta)\ \ \ ; \ \ \
0\leq n\in \mathbb{Z}\  .
\end{equation}
The dimensionless constant $\eta$ in (\ref{Eq35}) can be determined
by substituting the leading order expansion (see Eq. 6.1.34 of
\cite{Abram})
\begin{equation}\label{Eq36}
{{1}\over{\Gamma({1\over2}+\beta+ik-i\varpi)}}=(-1)^n
n!\eta\cdot\delta + O(\delta^2)\
\end{equation}
into the l.h.s of the resonance equation (\ref{Eq33}). This yields
\begin{equation}\label{Eq37}
\eta={{{\cal F}}\over{(-1)^n n!}}\  ,
\end{equation}
where [see Eq. (\ref{Eq33})]
\begin{equation}\label{Eq38}
{\cal F}=\Big[{{\Gamma(-2\beta)}\over{\Gamma(2\beta)}}\Big]^2 {{\Gamma({1\over 2}+\beta-ik)\Gamma({1\over
2}+\beta-\kappa)}\over{\Gamma({1\over
2}-\beta-ik)\Gamma({1\over 2}-\beta-\kappa)\Gamma({1\over
2}-\beta+ik-i\varpi)}}\  .
\end{equation}

\section{Summary}

In summary, in this paper we have studied {\it analytically} the
quasi-bound state resonances of charged massive scalar fields in the
near-extremal charged Reissner-Nordstr\"om black-hole spacetime.
These characteristic resonant modes of the composed black-hole-field
system are characterized by the physically motivated boundary
condition of purely ingoing waves at the black-hole horizon and
asymptotically decaying (bounded) radial eigenfunctions at
asymptotic spatial infinity.

Solving the Klein-Gordon wave equation for the linearized charged
massive scalar fields in the near-extremal ($MT_{\text{BH}}\to0$)
charged black-hole spacetime, we have derived the compact analytical
formula [see Eqs. (\ref{Eq9}), (\ref{Eq12}), (\ref{Eq13}),
(\ref{Eq14}), (\ref{Eq34}), and (\ref{Eq35})] \cite{Noteqqm}
\begin{equation}\label{Eq39}
\omega_n=\omega_{\text{c}}+2\pi T_{\text{BH}}\cdot
\big\{qQ-i\big[n+{1\over
2}+\sqrt{(l+1/2)^2+M^2(\mu^2-q^2)}\big]+O(MT_{\text{BH}},\delta)\big\}\
\ \ ; \ \ \ n=0,1,2,...
\end{equation}
for the complex resonant spectrum which characterizes the
quasi-bound state configurations of the composed
Reissner-Nordstr\"om-black-hole-charged-massive-scalar-field system.

\bigskip
\noindent
{\bf ACKNOWLEDGMENTS}
\bigskip

This research is supported by the Carmel Science Foundation. I thank
Yael Oren, Arbel M. Ongo, Ayelet B. Lata, and Alona B. Tea for
stimulating discussions.



\begin{thebibliography}{99}
\bibitem{Notewm} It is worth mentioning the physically interesting fact that,
as explicitly shown in \cite{Hodrc,Herkr},
non-spherically symmetric rotating black holes in asymptotically
flat spacetimes can support external stationary matter
configurations which are made of (neutral or charged) massive scalar
fields.
\bibitem{Notelf} In particular, as shown in \cite{Hercms}, the lifetimes of the
external quasi-bound matter configurations can be very large on the
scale $M$ set by the black-hole mass.
\bibitem{Noteqnm} It is worth noting that the familiar
quasinormal resonances (free oscillations) of the composed
Reissner-Nordstr\"om-black-hole-charged-massive-scalar-field system
studied in \cite{Hodch1,Hodch2,Konf,Ric} are characterized by the
boundary condition of free outgoing waves at spatial infinity. On
the other hand, as emphasized above, the quasi-bound state
black-hole-field resonances that we shall study in the present paper
are characterized by exponentially decaying (bounded) radial
eigenfunctions at spatial infinity [see Eqs. (\ref{Eq10}) and
(\ref{Eq11}) below] \cite{Hercms}.
\bibitem{Noteun} We shall use natural units in which $G=c=\hbar=1$.
\bibitem{Notemq} Here $M$ and $Q$ are respectively the mass and
electric charge of the central Reissner-Nordstr\"om black hole.
\bibitem{Noteqm} Note that $\mu$ and $q$ stand for $\mu/\hbar$ and $q/\hbar$, respectively.
Hence, these physical parameters of the charged massive scalar fields have the dimensions of
$($length$)^{-1}$.
\bibitem{Notegen} We shall assume $Q>0$ and $q>0$ for the composed black-hole-field system without loss of
generality.
\bibitem{Noteom} The integer parameters $m$ and $l\geq |m|$ are respectively the
azimuthal harmonic index and the spherical harmonic index which
characterize the scalar field mode. We shall henceforth omit these
harmonic indices for brevity.
\bibitem{Notell} Here $l(l+1)$ with $l=0,1,2,...$ are the familiar
discrete eigenvalues of the angular spherical harmonic functions
$S_{lm}(\theta)$.
\bibitem{Noteho} Note that the near-horizon limit $r\to r_+$ corresponds to the
asymptotic radial limit $y\to -\infty$.
\bibitem{Notenr} Here the integer $n$ is the resonance parameter which characterizes
the quasi-bound state resonances of the charged massive scalar
fields [see Eq. (\ref{Eq35}) below].
\bibitem{Notebi} It is worth mentioning that the case $i\beta\in
\mathbb{R}$ was studied in \cite{Hodch1}.
\bibitem{Notematc} Note that this matching procedure is valid in the
low-temperature $\tau\ll1$ regime of near-extremal black holes.
\bibitem{Notephr} That is, the mathematical solution of the radial equation (\ref{Eq15})
which respects the physically motivated boundary condition
(\ref{Eq8}) of purely ingoing waves at the outer black-hole horizon.
\bibitem{Noteqqm} Here we have used the fact that $k=qQ+O(MT_{\text{BH}})$
in the near-extremal $T_{\text{BH}}\to 0$ regime with
$\omega=\omega_{\text{c}}+O(T_{\text{BH}})$ [see Eqs. (\ref{Eq9})
and (\ref{Eq12})].
\end{thebibliography}
\end{document}